\documentclass[usenatbib]{mn2e}
\usepackage{graphicx}

\usepackage{amssymb}
\usepackage{color}
\usepackage{float}
\usepackage{amsmath}
\usepackage{appendix}
\usepackage{booktabs}
\usepackage{multirow}
\usepackage{siunitx}

\newcommand{\bn}{\begin{enumerate}}
\newcommand{\en}{\end{enumerate}}
\newcommand{\bi}{\begin{itemize}}
\newcommand{\ei}{\end{itemize}}

\def\gtorder{\mathrel{\raise.3ex\hbox{$>$}\mkern-14mu
    \lower0.6ex\hbox{$\sim$}}}
\def\ltorder{\mathrel{\raise.3ex\hbox{$<$}\mkern-14mu
    \lower0.6ex\hbox{$\sim$}}}

\setcitestyle{aysep={},citesep={,}} 

\newcommand{\rh}{r_{\rm h}}
\newcommand{\trhi}{t_{\rm rh,i}}
\newcommand{\deltam}{\Delta m}
\newcommand{\deltamz}{\deltam_{\rm z}}
\newcommand{\deltamx}{\deltam_{\rm x}}

\title[Central Dynamics of Rotating Star Clusters]{Central Dynamics of Multi-mass Rotating Star Clusters}
\author[M. Tiongco, A. Collier, and A.~L. Varri]
{Maria Tiongco$^{1}$\thanks{E-mail: maria.tiongco@colorado.edu}, 
Angela Collier$^{1}$,
and Anna Lisa Varri$^{2,3}$
\\
\footnotemark    
$^{1}$ JILA and Department of Astrophysical and Planetary Sciences, CU Boulder, Boulder, CO 80309, USA\\
$^{2}$Institute for Astronomy, University of Edinburgh, Royal Observatory, Blackford Hill, Edinburgh EH9 3HJ, UK\\
$^{3}$School of Mathematics, University of Edinburgh, Kings Buildings, Edinburgh EH9 3JZ, UK}

\begin{document}

\date{Accepted ?; Received ??; in original form ???}


\maketitle

\begin{abstract}
We investigate the evolutionary nexus between the morphology and internal kinematics of the central regions of collisional, rotating, multi-mass stellar systems, with special attention to the spatial characterisation of the process of mass segregation.  We report results from idealized, purely $N$-body simulations that show multi-mass, rotating, and spherical systems rapidly form an oblate, spheroidal massive core, unlike single-mass rotating or multi-mass non-rotating configurations with otherwise identical initial properties, indicating that this evolution is a result of the interplay between the presence of a mass spectrum and angular momentum. This feature appears to be long-lasting, preserving itself for several relaxation times.   The degree of flattening experienced by the systems is directly proportional to the initial degree of internal rotation. In addition, this morphological effect has a clear characterisation in terms of orbital architecture, as it lowers the inclination of the orbits of massive stars.  We offer an idealised dynamical interpretation that could explain the mechanism underpinning this effect and we highlight possible useful implications, from kinematic hysteresis to spatial distribution of dark remnants in dense stellar systems.
\end{abstract}



\section{Introduction}
\label{sec:intro}

The traditional picture of globular clusters as fully relaxed, isotropic, non-rotating, spherical systems characterised by a single old stellar population cannot last against the complexities emerging from the new-generation data which are now available and the theoretical ambitions that they stimulate. In particular, after a few pioneering efforts \citep[e.g., see][]{AndersonKing2003},  there is now a convincing body of observational investigations mapping the internal kinematics of several Galactic globular clusters \citep[e.g., see][]{Bianchini2013, Fabricius2014, Watkins2015, Ferraro2018, Kamann2018}. Recent astrometric studies based on Gaia Data Release 2 have further propelled the exploration of the degree of anisotropy in the three-dimensional velocity space \citep[e.g., see][]{Jindal2019}, and confirmed the growing evidence that the presence of internal rotation in globular clusters is much more common than previously assumed \citep[e.g., see][]{Bianchini2018, Sollima2019,Vasiliev2019}. 

This kinematic richness is now progressively being leveraged to attack a number of outstanding questions concerning the internal dynamics of this class of stellar systems, from the radial distribution of their angular momentum content \citep[e.g., see][]{Bellini2017, Lanzoni2018} to the phase space properties of their present-day stellar populations \citep[e.g., see][]{Richer2013, Cordero2017,Cordoni2020}.

Kinematic studies such as the those mentioned above find an essential counterpart in detailed investigations of the structural  and morphological properties of star clusters, which, after some early analyses \citep[e.g., see][]{Geyer1983,WhiteShawl1987,Kontizas1989}, unfortunately, remain relatively scarce \citep[e.g., see][]{ChenChen2010, Stetson2019}. The exploration of the natural connection between kinematics and morphology is indeed a crucial step to fully understand the intrinsic phase space structure of these stellar systems (e.g., see the informative studies conducted by \citealt{Davoust1990}, \citealt{Han1994}, \citealt{Ryden1996}, and, more recently \citealt{vandenBerg2008}). Such a joint approach can offer great insight into the physical origin of their angular momentum content (e.g., see \citealt{Frenk1982,Fall1985}), the importance of any tidal perturbation and, more generally, the different phases of their dynamical evolution, as driven by the synergy of internal and external processes.   

Theoretical and numerical studies have indeed shown that the total angular momentum content in collisional stellar systems is directly impacted by two-body relaxation processes which determine its redistribution, transport and eventual loss, especially in the case of tidally perturbed systems \citep[e.g., see][]{EinselSpurzem1999, Ernst2007, Hong2013, Tiongco2017}. Therefore, the strength and distribution of the angular momentum we measure in Galactic globular clusters today are remnant signatures of the initial rotation content imprinted in such systems by their formation processes (e.g., see \citealt{Lanzoni2018b} for a comparison between the present-day rotation curve of NGC 5904 and a long-term N-body simulation from the survey by \citealt{Tiongco2016}).

More generally, even in the case of an isolated system, the presence of non-vanishing total angular momentum may lead to a more complex long-term dynamical evolution compared to the one of a non-rotating system (e.g., the suggestion by \citealt{hachisu1979} of the existence of a ``gravo-gyro catastrophe", subsequently explored by several other authors). Additional investigations have also highlighted interesting effects of the interplay between bulk internal rotation and a mass spectrum of stars in a stellar system. In particular, \citealt{kim2004} showed that systems of that kind can produce an oblate core of fast rotating heavy masses, similar to the one identified in the present study. Most recently, \citet{sz2019} showed that in rotating star clusters, the orbital inclinations of the heaviest stars decrease over time, creating a mass segregation effect in the distribution of orbital inclinations in addition to the well-known radial (isotropic) mass segregation effect.  This effect has also been observed in simulations of stellar systems orbiting a massive black hole, such as nuclear star cluster simulations of \citet{sz2018}, the eccentric nuclear disk simulations of \citet{foote2019}.

So far, the explanation of why this effect occurs has been attributed to resonant relaxation and resonant friction, which was first introduced by \citet{rt96}. In a stellar system, coherent torques from stars on stable orbits enhance the rate of angular momentum relaxation.  In a stellar system dominated by a central mass, both the magnitude and direction of the angular momentum vectors change in a random walk fashion, while in a stellar system without a dominant central mass, only the direction of the angular momentum vectors change stochastically.  The latter process is referred to as vector resonant relaxation and is considered a more limited form of resonant relaxation.  \citet{meiron2019} studied the effects of vector resonant relaxation in globular cluster like systems.  \citet{rt96} also coined the term ``resonant friction'' that describes how the orbital inclinations of massive objects in a stellar system can be lowered by near-resonances. An application of such concepts to the study of the statistical mechanics of rotating systems with a central black hole has been recently presented by \citet{Gruzinov2020}.

In this work, we wish to concentrate on the investigation of the evolutionary nexus between morphology and kinematics of the central regions of collisional, rotating systems, with special attention to the process of mass segregation. We perform and interpret a new series of $N$-body simulations of rotating globular clusters with a spectrum of stellar masses. In view of our specific interest in the central dynamics and structural properties, we restrict our investigation to a set of initial conditions characterised by spherical symmetry that distinguishes our study from previous similar investigations that used initially oblate, i.e., already flattened, rotating models (commonly used models include the ones from \citealt{lupton1987} and \citealt{varri2012}).  Our parameter space explores different degrees of rotation and velocity anisotropy, and we provide evidence that the latter also plays a non-trivial role in the development of the so-called `anisotropic mass segregation'.  We explore in depth such mass segregation process along different spatial directions within the cluster, and note that the central regions of the cluster are flattened as an oblate spheroid that develops and sustains itself for several relaxation times.  We also offer an idealised dynamical interpretation that could explain the mechanism underpinning these effects and we present a discussion of possible implication our results.

This article is structured as follows: numerical aspects and initial conditions are described in Section \ref{sec:numerics} and our results from numerical modeling are presented in Section \ref{sec:results}. Next, we discuss some analytical aspects of our results with some additional numerical experiments to understand this analysis further in Section \ref{sec:discussion}. Conclusions drawn from this work are reported in Section \ref{sec:conclusion}.

\begin{table}
\caption{Summary of the properties of the $N$-body models presented in this study (in H\'enon units). The 2nd column shows the ratio of the initial rotational kinetic energy and total kinetic energy in the system, the 3rd column reports the initial value of the spin parameter (Eq. \ref{eq:spin}) for each model, and the final two columns denote the fraction of prograde particles for each mass bin (see Section 2 for definitions). Rows 1-4 correspond to the primary set of $N$-body models ordered by increasing spin; rows 5-9 refer to the additional $N$-body experiments. }
  \begin{center}
  \begin{tabular}{lllllSS}
    \toprule
    \multirow{2}{*}{Model} & \multirow{2}{*}{$R_E/R_{KE}$} & \multirow{2}{*}{$\lambda$}  & 
     \multicolumn{2}{c}{Prograde Fraction (\%)} \\
     & & & {H} & {L} \\
      \midrule
    $R000$ & 0.001 &0.001 &50 &50  \\
    $R050$  & 0.026 &0.058  &75 &75 \\
    $R075$  & 0.056 &0.084   &87.5 &87.5 \\
    $R100$  & 0.101 &0.110   &100 &100\\
    $R_{h}$  & 0.045 &0.085    &100 &50 \\
    $R_{l}$  & 0.011 &0.028  &50 &100 \\
    $R100S50$  & 0.131 &0.111   &100 &100\\
    $R100S75$  & 0.165 &0.112 &100 &100\\
    $R_{r}$  & 0.084 & 0.103 & 88.8 & 88.8\\
    \bottomrule
  \end{tabular}
  \end{center}
\label{table:1} 
\end{table}

\section{Method and Initial Conditions}
\label{sec:numerics}

In our survey of $N$-body simulations, the initial conditions are first  defined by a \citet{king1966} spherical, isotropic, non-rotating distribution function, with the concentration parameter chosen to be $W_0=6$. For the mass spectrum,  we adopted a power-law distribution with a slope of -2 and the mass ratio of the heaviest star to the lightest star set to 100 (the resulting ratio of the heaviest star mass to the average stellar mass is $\approx$ 21). Each $N$-body model has a total of $N=$ 65~536 particles. The initial conditions were generated using the McLuster code \citep{kuepper2011}.

Rotation is introduced into the system by randomly selecting a fraction of particles and setting their tangential velocities in the same direction, i.e., their tangential velocities are set as their absolute value. This implementation preserves the solution of the Boltzmann equation and also the shape of the chosen equilibria; we refer to this change in the velocities as the action of ``Lynden-Bell's demon'' (\citealt{lynd60}, see also \citealt{Rozier2019} for a recent application). We note that the phase space invariance under velocity reversals is a direct result of the Jeans theorem \citep{jean19}. The models are named $R$ for ``rotating'', followed by a number that denotes the fraction of particles that have been selected for the application of the ``demon''; we use the terms ``prograde" and ``retrograde" to mean that $v_{\phi}>0$ or $<0$, respectively, and the coordinate system is such that  the axis of rotation corresponds to the $z$ direction).
For example, the $R100$ model has $100\%$ of particles rotating in the same direction. The models $R000$, $R050$, $R075$, and $R100$ comprise the primary models of our study, in order of increasing rotation, and have identical initial positions of stars.

We measure the angular momentum content of each $N$-body model via two distinct metrics. First, we calculate the kinetic energy due to rotation and compare to the total kinetic energy in the model. The second parameter measured to understand spin is the well-known ``spin parameter'' (e.g., \citealt{peeb69}), 

\begin{eqnarray}
\label{eq:spin}
\lambda=\frac{JE^{1/2}}{GM^{5/2}}
\end{eqnarray}

\noindent
where $J$, $E$, and $M$ are the total angular momentum, energy and mass respectively, and $G$ is the gravitational constant.  Throughout the paper, we adopted the H\'enon  system of units \citep{henon1971,heggie1986}, where $G=M=1$ and $E=-0.25$. These two measures of angular momentum content are listed in Table \ref{table:1}.  The fiducial, isotropic, $R000$ model has an initial $\lambda$ of $\sim 0$, and we increase $\lambda$ by increasing the fraction of prograde orbits.

We have also run some additional experiments to complement our results, which we will describe in detail in Section 4.  To understand the roles of the different mass components, we have created some additional $N$-body models by dividing the given initial equilibrium into mass bins and by rotating the bins individually (Experiment 1).  The most massive $1/3$ of particles are denoted as ``high mass'' (H) and the rest of the particles (the lower $2/3$) are marked as ``low mass'' particles (L).

In addition, we have also considered a subset of initial conditions (Experiment 2) characterised by some degree of isotropic/radial mass segregation, as there is dynamical and observational evidence suggesting the existence of primordial mass segregation in young star clusters \citep[e.g., see][]{Bonnell1998,deGrijs2002,McMillan2007}.  The prescription adopted to introduce such mass segregation is based on the one featured in \citet{baumgardt2008}, where, in short, by setting the segregation parameter $S$ closer to 1, the more likely a heavy particle is initialized closer to the centre, with a segregation parameter of 1 being fully segregated (with the heaviest particle at the shortest radius from the centre, followed by the next heaviest particle at the 2nd shortest radius, etc.).

\begin{figure*}
\centerline{
\includegraphics{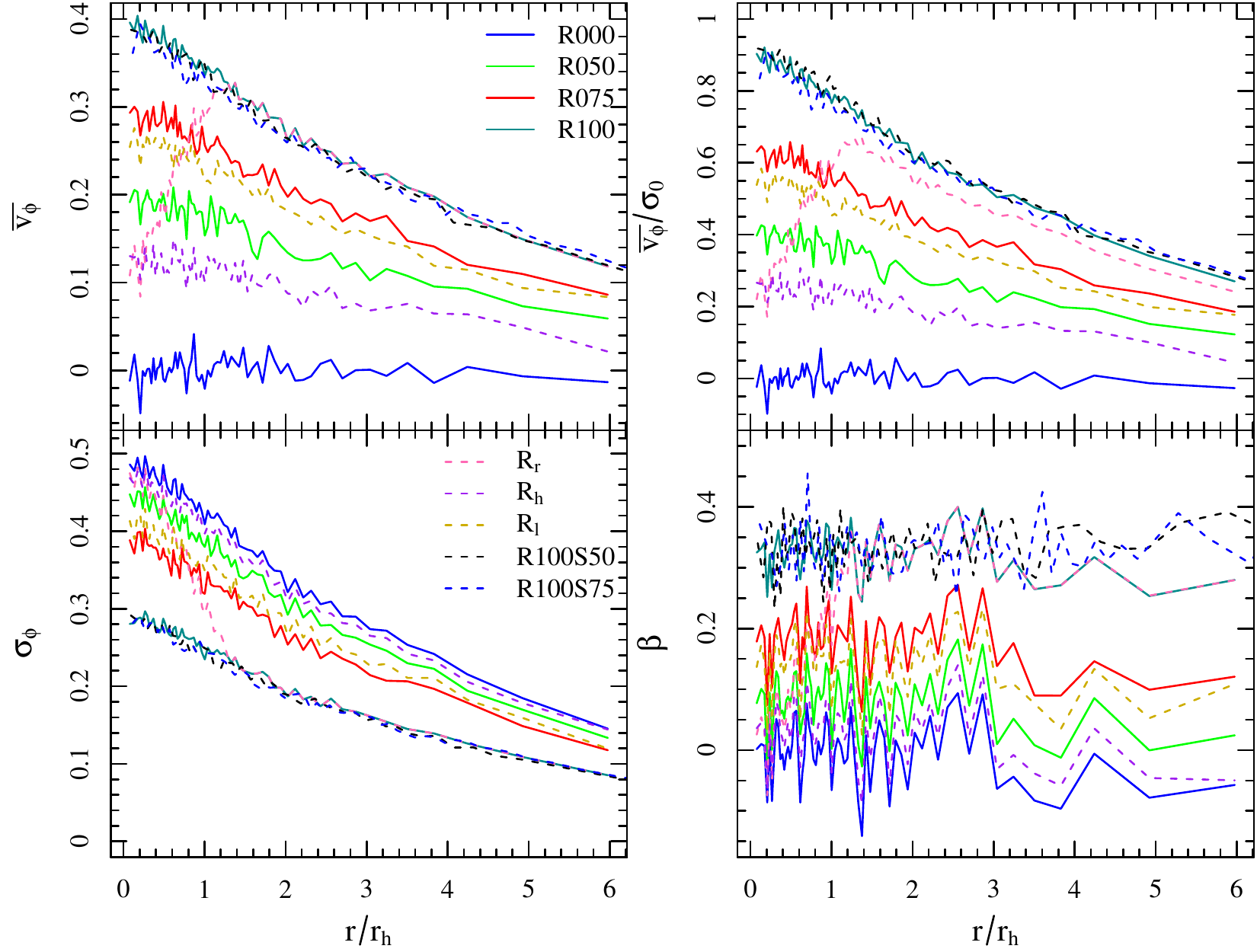}}
\caption{Top left: Initial rotational profile (mean velocity in the azimuthal/$\phi$ direction), as a function of cylindrical radius normalized to the projected half-mass radius.  Bottom left: Initial velocity dispersion $\sigma_{\phi}$ profile. Top right: Initial rotational profile normalized to the central velocity dispersion $\sigma_0 = \sqrt{\frac{1}{3}(\sigma_{\rm r,0}^2 + \sigma_{\rm \phi,0}^2 + \sigma_{\rm \theta,0}^2)}$. Bottom right: Initial velocity anisotropy parameter ($\beta$, see Eq. \ref{eq:beta}). In all panels, solid lines denote the primary $N$-body models (Table 1, Row 1-4). Dashed lines represent the additional N-body models discussed in Section \ref{sec:discussion} (Table 1, Row 5-9).}
\label{fig:ics}
\end{figure*}

Finally, we consider a more realistic rotation curve (Experiment 3) defined by the rotation curve increasing from zero from the centre of the cluster, peaking at approximately the half-mass radius, then decreasing further out \citep[see, e.g.,][]{Lanzoni2018,Tiongco2017}. This rotational profile was generated by reversing the tangential velocities of a different percentage of particles in each radial bin until the desired profile is achieved as shown in Figure \ref{fig:ics}.
Our survey of $N$-body simulations was performed using NBODY6 \citep{aarseth2003} with GPU acceleration \citep{nitadori2012}. All $N$-body models are evolved in isolation and the effects of stellar evolution are not included. 

A reference time scale that we have adopted in all our analyses is the initial half-mass relaxation time defined as
\begin{equation}
\trhi=\frac{0.138 N^{1/2}r_{\rm h}^{3/2}}{\langle m\rangle^{1/2} G^{1/2} \log(0.11 N)}
\end{equation}
where $\langle m \rangle $ is the mean stellar mass, and $r_{\rm h}$ is the 3D half-mass radius, the radius enclosing half the mass of the cluster (see e.g. \citealt{heggie2003}).  In H\'enon units, the $\trhi$ of all of our models is $\approx$ 730 time units.  For all of the models featured in this study, the duration of the simulations is of 4000 time units.

We show relevant kinematic properties as a function of radius for all of our models in Fig. \ref{fig:ics}, including rotational velocity, velocity dispersion, and velocity dispersion anisotropy; the importance of the latter two properties are discussed in the next section.  We acknowledge here that while introducing rotation into the system via Lyden-Bell's demon is straightforward, the resulting \textit{initial} rotation curves do not resemble what is observed in globular clusters.  However, as these systems evolve over a short period of time (i.e., over several dynamical times and much shorter than a relaxation time), their angular momentum distributions evolve into rotation curves which are comparable to the ones observed in present-day star clusters.  We show the same kinematic properties in Fig. \ref{fig:ics} after some short evolution in Fig. \ref{fig:t30}.  Overall, our main result shows how increasing the amount of angular momentum affects the spatial distribution of multi-mass stellar systems using initial conditions that \textit{do not change in physical structure when increasing the amount of rotation}, in contrast to initial conditions realized from commonly used distribution function based models with rotation, such as those from \citet{lupton1987} and \citet{varri2012}.

\section{Analysis of Primary Models}
\label{sec:results}

\subsection{Central morphology}
By the action of the ``Lynden-Bell's demon'', we have created $N$-body models with initial conditions that differ in their initial degree of rotation but maintain the original density and velocity distributions (i.e., their zeroth and second-order velocity moments). Therefore, the difference between the various $N$-body models appears  in their first-order velocity moment and corresponding velocity dispersion. 
 
For clarity, we assume the following definition of the velocity dispersion tensor \citep{binn08} 
\begin{eqnarray}
\sigma_{i,j}^2=\langle (v_i-\langle v_i \rangle)(v_j - \langle v_j \rangle)\rangle 
\label{eq:dispersion}
\end{eqnarray}

\noindent
where $i,j,k=r,\theta,\phi$ refer to conventional spherical coordinates.  While the initially non-rotating model (R000) is isotropic, the process of introducing rotation lowers the component $\sigma_{\phi}$, which, in turn, increases the associated degree of velocity anisotropy. We depict the initial $\sigma_{\phi}$ for each $N$-body model in the top panel of Fig. \ref{fig:ics}. The other components of the velocity dispersion tensor, $\sigma_r$ and $\sigma_{\theta}$, remain unchanged under the action of the ``Lynden-Bell's  demon'' and are identical to the $\sigma_{\phi}$ of the $R000$ model.  The bottom panel of Fig. \ref{fig:ics} illustrates the initial anisotropy parameter ($\beta(r)$) for each $N$-body model; the definition adopted here is
 
 \begin{eqnarray}
\beta=1-\frac{\sigma^2_{jj}+\sigma^2_{kk}}{2\sigma^2_{ii}}.
\label{eq:beta}
\end{eqnarray}

 \begin{figure*}
\centerline{
 \includegraphics[width=1.0\textwidth] {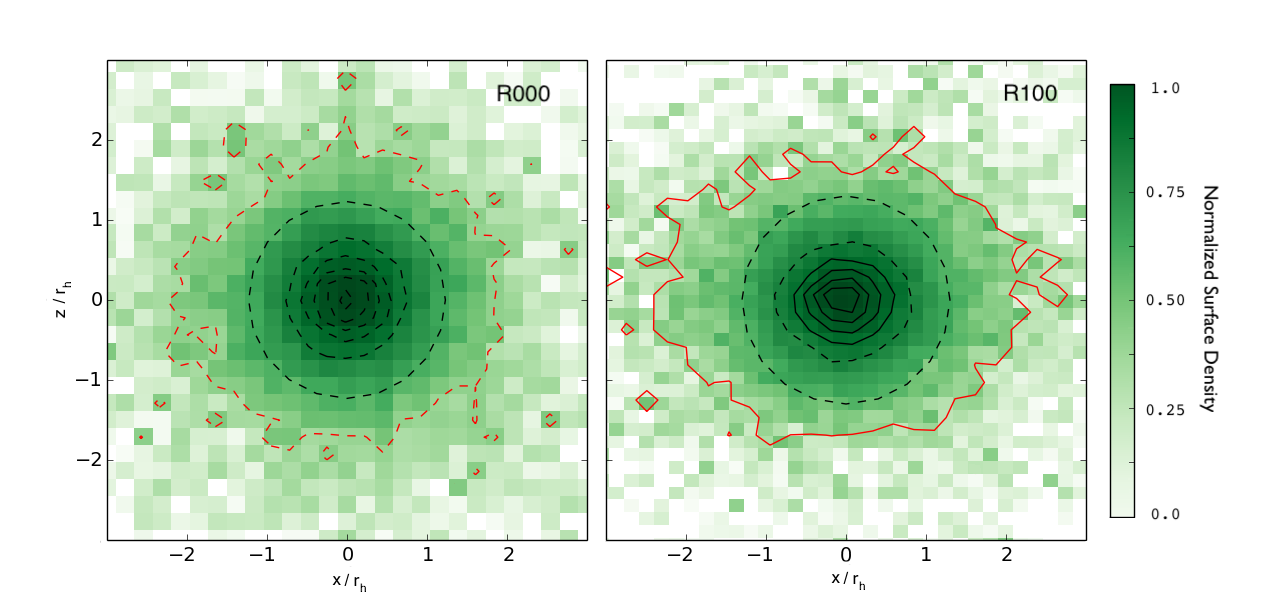}}
\caption{Projected isodensity contours and the surface density map projected onto the (x,z) plane of the $R000$ (left) and $R100$ (right) models. Solid lines indicate contours that are approximated as ellipses, while solid lines represent circular geometry. The outer contour (red) is the same level for both models and contains 90\% of the mass of the model. This figure is resulting from an average of multiple snapshots of the evolution of the $N$-body models around $t=1.38\trhi$. }
\label{fig:contours}
\end{figure*}

 \begin{figure}
\centerline{
 \includegraphics[width=0.5\textwidth] {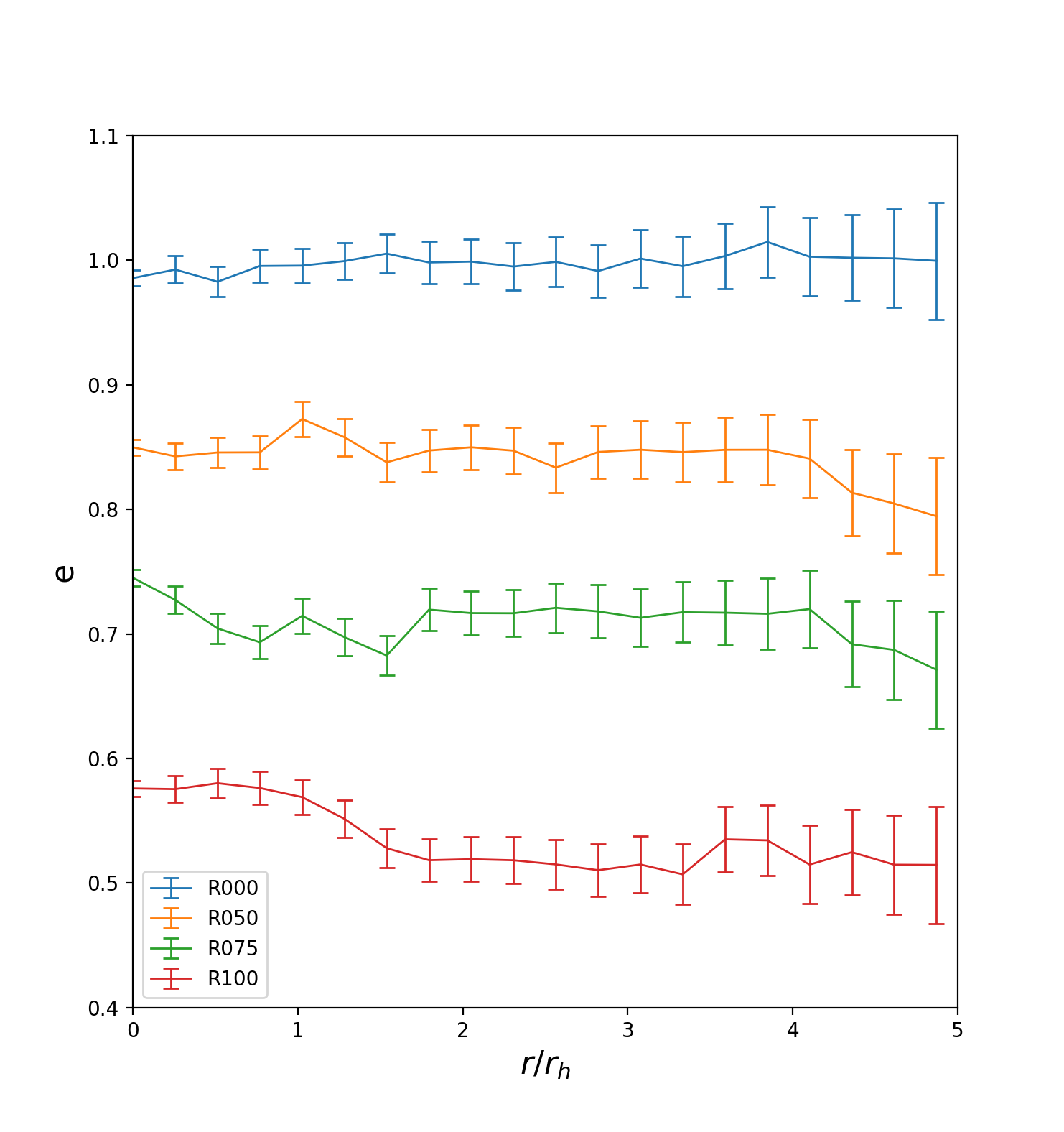}}
\caption{Ellipticity (defined as the ratio between the minor and the major axis) of the projected isodensity contours of all models in the $(x,z)$ plane at $t=1.38 \trhi$. The error bars represent the standard error of the mean for each bin.}
\label{fig:ellipticity}
\end{figure}

 \begin{figure}
\centerline{
 \includegraphics[width=0.5\textwidth] {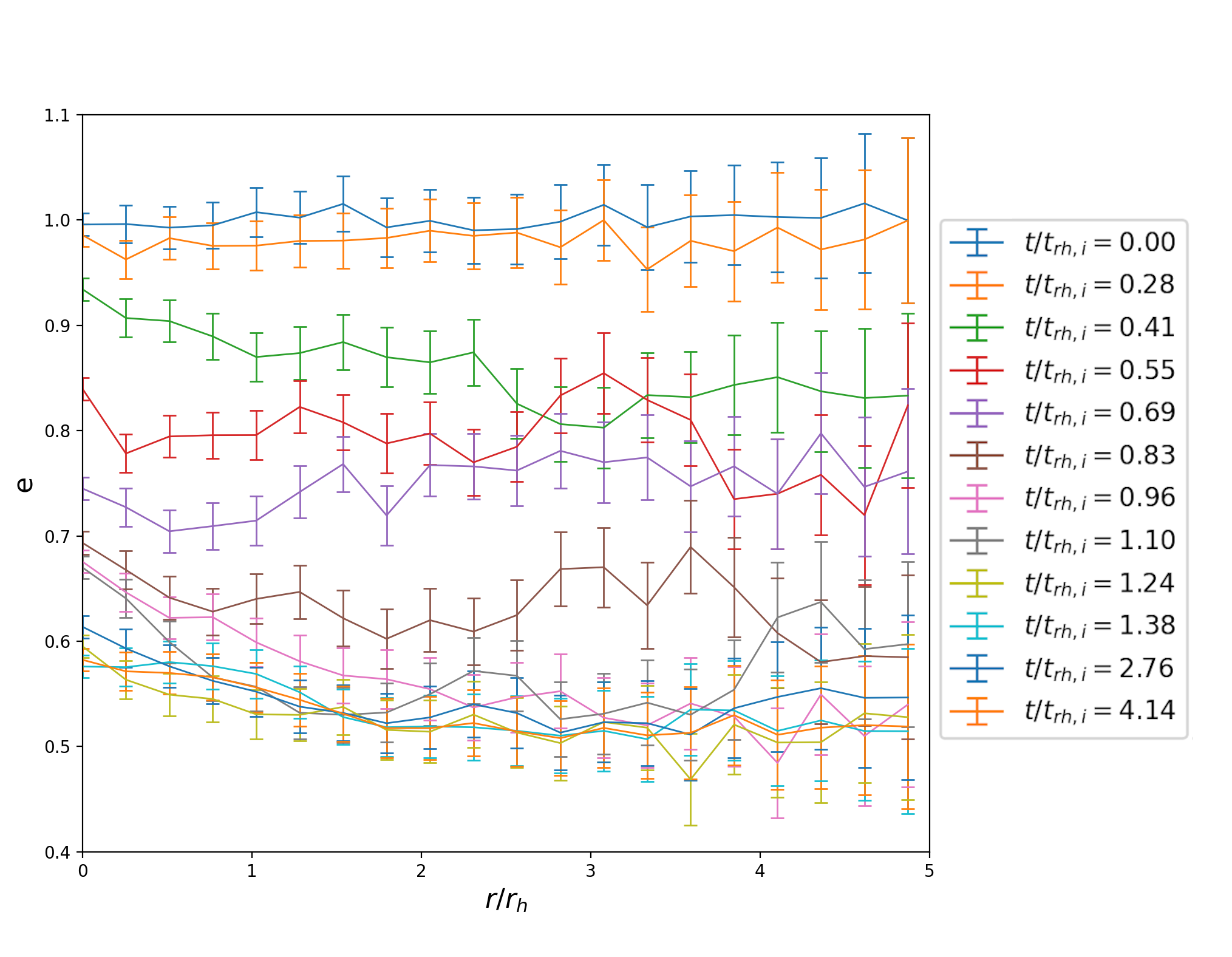}}
\caption{Ellipticity (defined as the ratio between the minor and the major axis) of the projected isodensity contours of the R100 model in the $(x,z)$ plane as it evolves in time.  The error bars represent the standard error of the mean for each bin.}
\label{fig:time_evo}
\end{figure}

 \begin{figure}
\centerline{
 \includegraphics{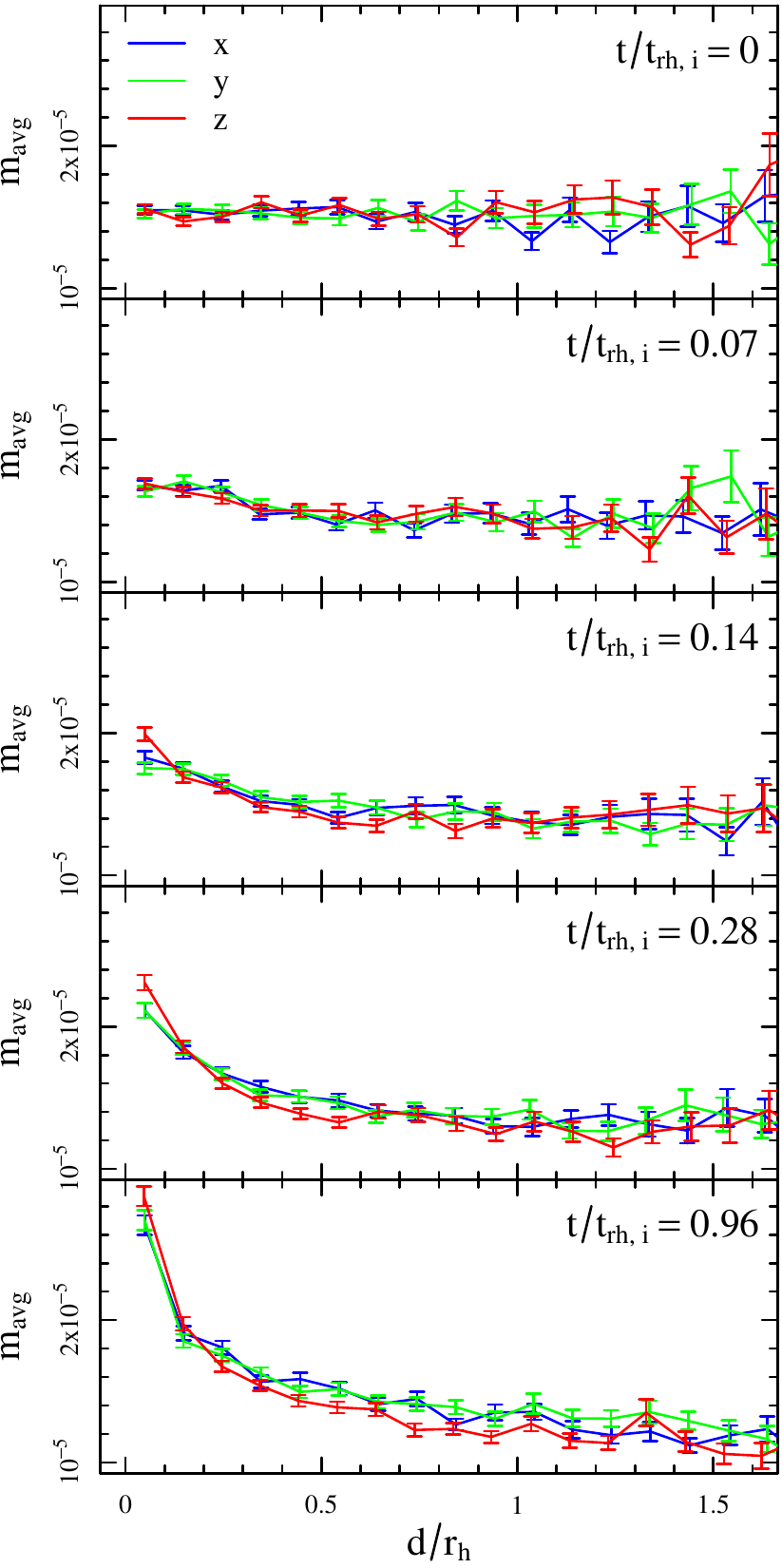}}
\caption{Evolution of the average stellar mass in a bin, represented as a function distance from the centre of the cluster, normalized to the half-mass radius, $\rh$, for model R100.  The bins are aligned along a system of reference centred in the centre of mass and such that the $z$-axis corresponds to the rotation axis and the $x$- and the $y$-axis are located within the plane perpendicular to it.  The error bars represent the standard error of the mean for each bin. The mass segregation parameter we use in this study is defined by the \textit{difference of the average mass} at the centre of the cluster and at the half-mass radius.}

\label{fig:msegprofiles}
\end{figure}

 \begin{figure}
\centerline{
 \includegraphics{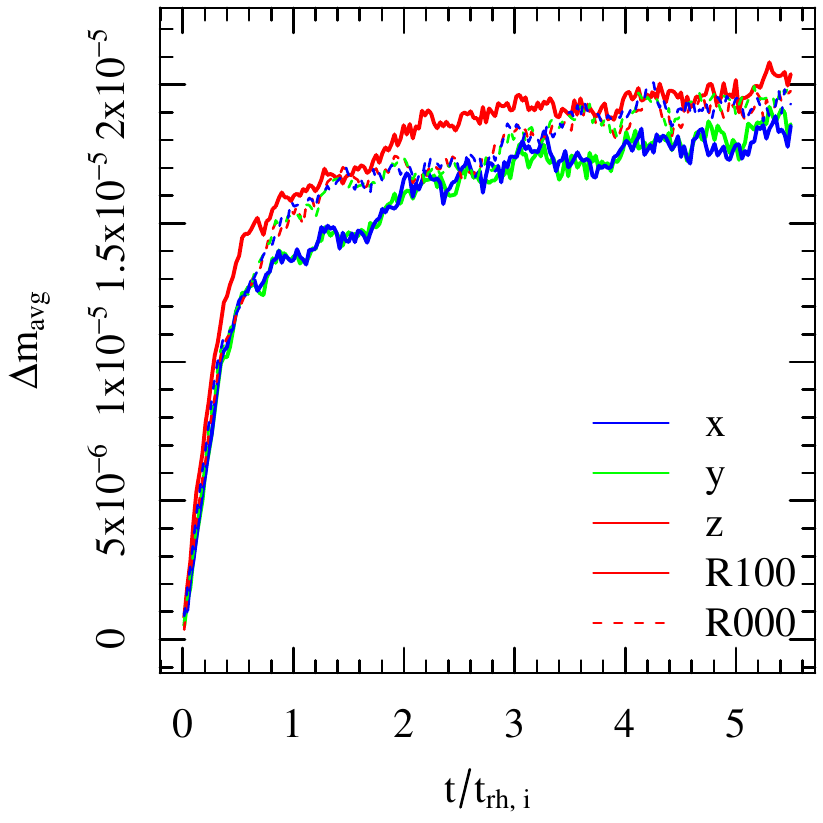}}
\caption{Evolution of $\deltam$ (see Eq. \ref{deltam} for definition) along the $x$, $y$, and $z$-axes, comparing the fully rotating model (R100) to the non-rotating model (R0).}
\label{fig:zmseg}
\end{figure}

The initial differences in the amount of rotation and degree of velocity anisotropy characterising our $N$-body models will subsequently determine, during their evolution, the degree of flattening acquired by the central regions of the systems.  We stress that the flattening seen in our $N$-body models should not be compared to the flattening observed in present-day rapidly rotating systems, such as certain classes of early-type galaxies, as the total amount of angular momentum introduced via ``Lynden-Bell 's demon'' is much smaller than the one possessed by an oblate spheroid whose morphology is completely shaped by rotation \citep[e.g., see][]{chandra,bin78}. 

In all of our rotating cases, we observe the appearance of a \textit{long-lasting oblate spheroidal structure in the core of the $N$-body model}, while, at larger radii, the system maintains a more spherical distribution. 
We illustrate this result in Fig. \ref{fig:contours}, showing the two-dimensional surface density maps and a selection of demonstrative isodensity contours of models $R000$ and $R100$ evaluated at 1.38$\trhi$ (i.e. a moment representative of a phase  long after the anisotropic mass segregation has peaked, discussed in the next section.) At this time in the simulation, the $R100$ model shows a central oblate shape in the $(x,z)$ plane, while the $R000$ model remains spherical throughout. The solid lines represent contours that can be approximated by ellipses while the dashed lines show a circular geometry; the contour containing 90\% of the total mass is denoted in red. We also note that rotating model has an oblate shape at this outer radii, while the non-rotating model is more spherical. 

To compare the morphology of all our $N$-body models, we have measured the ellipticity of isodensity contours of the each system again around 1.38$\trhi$, and plot the ellipticity as a function of radius in Fig.  \ref{fig:ellipticity}. In order to get a measurement for the smooth potential, we took the average of 15 time steps between 1.37 and 1.39$\trhi$ and measured the ellipticity profile of bins containing equal particle number. This figure shows that for the primary set of models: $R000$, $R050$, $R075$, and $R100$, that have increasing rotation along the sequence, the central regions of the system become flatter with faster rotation.  We will discuss this trend and the other models in Section \ref{sec:discussion}.

Finally, we use the same method of measuring ellipticity as a function of radius described in the previous paragraph to show how quickly our system evolves from the initially spherical structure to its final ellipticity profile. The ellipticity evolution of the $R100$ model is mapped in Figure \ref{fig:time_evo}. Within approximately one half-mass relaxation time, this model goes from spherical to increasingly oblate until equilibrium is reached and the model reaches an ellipticity profile that does not change much over a few relaxation times close to the end of the simulation.

\subsection{Anisotropic Mass Segregation and Changes in Orbitial Inclination}

Next, we seek to characterise the strength and spatial properties of the process of mass segregation occurring during the evolution of the primary $N$-body models considered in this study. We analyse the time evolution of the average stellar mass of the models, computed in bins defined along different directions within the systems. In such profiles, the particles are binned according to their $|x|$, $|y|$, or $|z|$ position in linearly spaced bins along the desired axis (the shape of the bins are parallel thin slabs with heights encompassing the entire cluster with the long sides parallel to the $x$-, $y$-, or $z$-axes, and two slabs for positive and negative positions). In Fig.   \ref{fig:msegprofiles} we illustrate this analysis for the representative case of model $R100$.

The presence of mass segregation can be recognised by the fact that the average mass of the innermost bin is higher compared to the rest of the profile. Such a feature means that the heaviest stars have migrated to the centre of the system, which also induces a further lowering of the average mass in the outer regions.  We compare the average mass profile along the $z$-axis (the rotation axis), with the profiles computed along the $x$- and the $y$-axes, which are expected to be indistinguishable because the system, while breaking its initial spherical symmetry (see previous section), remains axisymmetric throughout its evolution.  Initially, the $R100$ model does not show any primordial mass segregation by construction, but, as such a feature eventually emerges, the mass segregation along the $z$-axis becomes stronger than along the $x$- and $y$-axes.  This is indicated by the profile for the $z$-axis being higher than the other two axes in the central regions, and lower at some radius farther out, e.g., at the half-mass radius.

We now define a mass segregation parameter, 

\begin{equation}
\label{deltam}
    \deltam=m_{\rm avg,0} - m_{\rm avg,\rh}
\end{equation}

\noindent
where the first term on the right side is the average particle mass at centre of the system, and the second term is average particle mass at the half-mass radius (similar diagnostic tools have been adopted also by \citealt{gill2008,Trenti2013,Bianchini2016,Parker2016}, among others).  We will use this parameter to compare the evolution of the difference between the mass segregation along the $z$-axis versus the mass segregation along the $x$/$y$-axis for all of our $N$-body models throughout the paper.  We first demonstrate the use of $\deltam$ in Fig. \ref{fig:zmseg}, by comparing the time evolution of such an observable calculated along the $x$, $y$, and $z$-axes and plotted for models $R100$ and $R000$.  This analysis clearly shows that, in model $R100$, $\deltam$ along the $z$-axis becomes higher than along the other two axes (higher mass segregation along one direction, i.e., anisotropic mass segregation), while in model $R000$, $\deltam$ is the same along all three axis (equal amount of mass segregation along all directions, i.e., isotropic mass segregation).

\begin{figure}
\centerline{
 \includegraphics{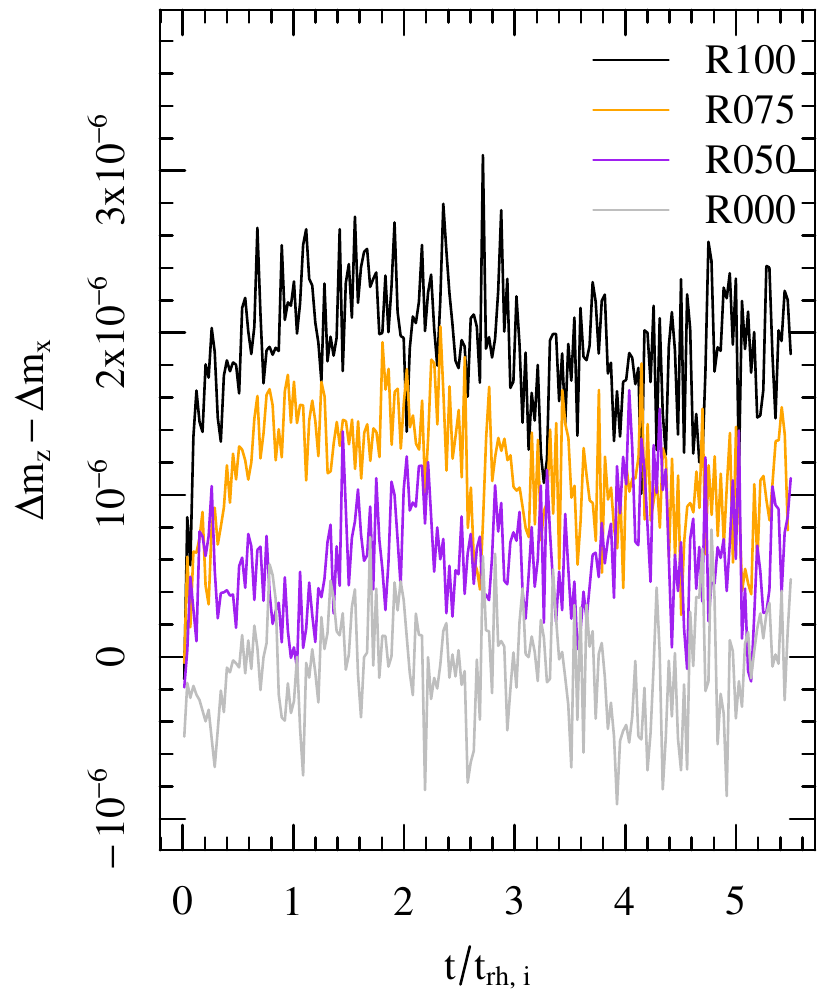}}
\caption{Evolution of the difference between the $\deltam$ parameter (see Eq. \ref{deltam} for definition) measured on the $z$-axis and on the $x$-axis, for all primary $N$-body models in our survey, from fully rotating to non-rotating.}
\label{fig:rotcompare1}
\end{figure}

Next, we compare the models $R100$, $R075$, $R050$, and $R000$, which go from fully-rotating to non-rotating.  For convenience, in Fig. \ref{fig:rotcompare1},  instead of examining $\deltam$ along the three Cartesian coordinates, we now directly illustrate the evolution of the difference between $\deltam$ along the $z$-axis and $\deltam$ along the $x$-axis for each model.  Again, we find that such a difference is more prominent in the case of models characterised by a higher degree of initial rotation, and agrees with the observation that these models are more flattened in the center; we will discuss the origin of this feature in Section \ref{sec:discussion}.  Finally, we find that the flattened morphology which develops in the central regions persists, for nearly all the cases, for at least several relaxation times (see Figs \ref{fig:zmseg} and \ref{fig:rotcompare1}).  Notable as well is that such an effect is not affected by the process of core-collapse of the systems, which occurs at approximately 0.6 $\trhi$.

\begin{figure}
\centerline{
 \includegraphics{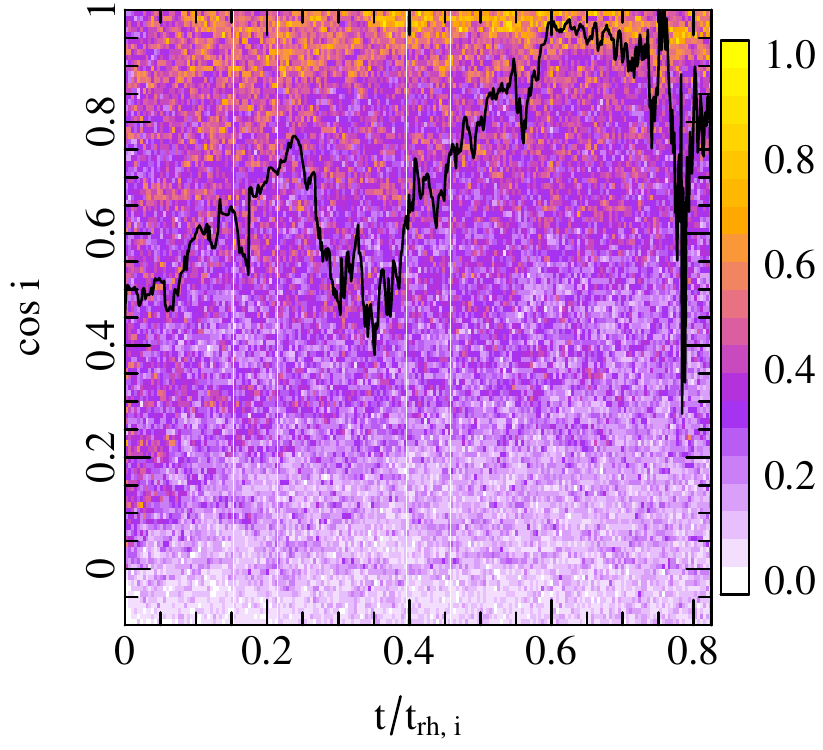}}
\caption{Inclination ($\cos{i}=L_{\rm z}/|\mathbfit{L}|$) of the one-percent heaviest stars orbits versus time in model $R100$ (see text for explanation).  The colours represent the density of points in the figure normalized to the highest density.  The solid black line shows the evolution of one star to demonstrate the evolution towards a lower orbital inclination.}
\label{fig:orbitinc}
\end{figure}

We now wish to investigate the orbital architecture associated with the morphological and dynamical properties identified in the previous two sections. To form a long lasting oblate structure in the core of the systems, the inclination of the orbits in the $(x,z)$ plane must decrease. We show the evolution of the orbital inclinations of the one-percent heaviest particles in Fig. \ref{fig:orbitinc} for model $R100$.  The orbital inclination is defined to be $\cos{i}=L_{\rm z}/|\mathbfit{L}|$, where $\mathbfit{L}$ is the angular momentum vector of the stellar orbit.  Due to the overlapping of hundreds of lines, we use a colour map to represent the density of the points in Fig. \ref{fig:orbitinc}, along with a solid black showing the evolution of the inclination of one particle that illustrates the lowering of its orbital inclination.  The figure shows that, in the beginning of the simulation, the inclinations of the heaviest particles are distributed uniformly, but, after some evolution, there are more lower inclination orbits (where $\cos{i}=1$ would be the lowest) than higher inclination orbits among the heaviest particles.

\section{Dynamical Interpretation and Additional Experiments}
\label{sec:discussion}
We have analysed the long-term evolution of collisional, multi-mass, initially spherical N-body models with varying amounts of bulk internal rotation, and we have noted the emergence and persistence of an oblate spheroid in the central regions of the systems. Based on these results, we offer here an interpretation of this long-term dynamical behavior, with emphasis on the role played by the angular momentum. Next, we test this interpretation with additional numerical experiments.

\subsection{The Role of {\it {\textbf{Velocity}}} Anisotropy}
The rotation-induced oblateness of astronomical bodies is a classical problem in Newtonian and celestial mechanics (e.g., see \citealt{chand68,chandra}, \citealt{bin78} and most recently \citealt{kir19}). The rotational movement of a solid body gives rise to centrifugal accelerations which eventually leads to deformation of the body itself. As discussed in Section \ref{sec:results} and depicted in Fig. \ref{fig:ics}, the fiducial models ($R000$, $R050$, $R075$, and $R100$) vary only in initial velocity dispersion and anisotropy profile. When spinning up the initial configuration by means of the ``Lynden-Bell's demon'', we decrease only $\sigma_{\phi}$, which allows us to use the following relation between the components of the velocity dispersion tensor,
\begin{eqnarray}
\sigma_{ii}=\sigma_{jj}=\frac{\sigma_{ij}}{\sqrt{2}}.
\end{eqnarray}

We can then rewrite the velocity dispersion from Eq. \ref{eq:dispersion} in terms of $\beta$, which is uniquely determined for each model, and $\sigma_{ij}$, which does not change for our system:

\begin{eqnarray}
\sigma^2=\frac{1}{2}\sigma^2_{ij}(3-\beta).
\label{eq:final_eq}
\end{eqnarray}

The non-rotating model $R000$ is supported against gravitational collapse by the random motions of the particles.  By introducing rotation in our initial conditions by means of the ``Lynden-Bell's demon'', we have also increased $\beta$ (see Fig.~1), which, in turn, causes a collapse along the direction of rotation by lowering the velocity dispersion components according to Eq. \ref{eq:final_eq}. Such an effect can be physically interpreted as a deviation from a condition of hydrostatic equilibrium, with attention to the role played by the pressure gradient.  The resulting variation in the pressure gradient scales as  $-\rho(r)\sigma^2$, where $\rho(r)$ is the volume density \citep[see, e.g.,][]{binn08}. The changing pressure gradient requires a particle to experience a net pressure force acting toward the centre of the system. The flattening is not seen in the $x$ and $y$ directions due to the appearance of the centrifugal force.

In the following sections, we examine the effect of increased velocity anisotropy creating a negative pressure gradient and flattening the rotating core of the cluster. The presence of a mass spectrum is discussed in \citet{sz2019} as a necessary and sufficient condition to produce this flattening effect. We note that we have independently confirmed this statement by performing also an additional $N$-body simulation of a single-mass rotating system (not included here for brevity, see also \citet{meza2002} for more in-depth exploration of similar systems), and we found no obvious flattening of the core. This implies that, while velocity anisotropy is important for this effect, the radial mass segregation found in multi-mass models must also play a role.   

In order to understand the relative importance of velocity anisotropy and mass segregation we have performed additional experiments that we will describe in greater detail below. First, we analyse $N$-body models where we selectively rotate only high mass particles (model $R_h$) or only low and intermediate mass particles (model $R_l$). Next, we present and analyse two $N$-body models characterised by the same amount of initial rotation simulation  and different degree of primordial mass segregation: models $R100S50$ and $R100S75$.

\subsubsection{Experiment 1: Selective Rotation by Mass}
We start from the same initial equilibria described in Section 2 and we introduce some non-vanishing angular momentum exclusively in the heavy component of the system, by forcing $100\%$ of the $1/3$ most massive particles to rotate in the same direction. The velocity vectors of the remaining particles are untouched, and their distribution is isotropic. 
The properties of these initial conditions are illustrated in Fig. \ref{fig:ics} and the system is labeled as $R_h$. 
The velocity anisotropy profile of this model lies in between models $R000$ and $R050$.
If the flattening effect depends only on velocity anisotropy, then we would expect the measurement of $\Delta m$ to show little flattening as well. 

 \begin{figure}
\centerline{
 \includegraphics{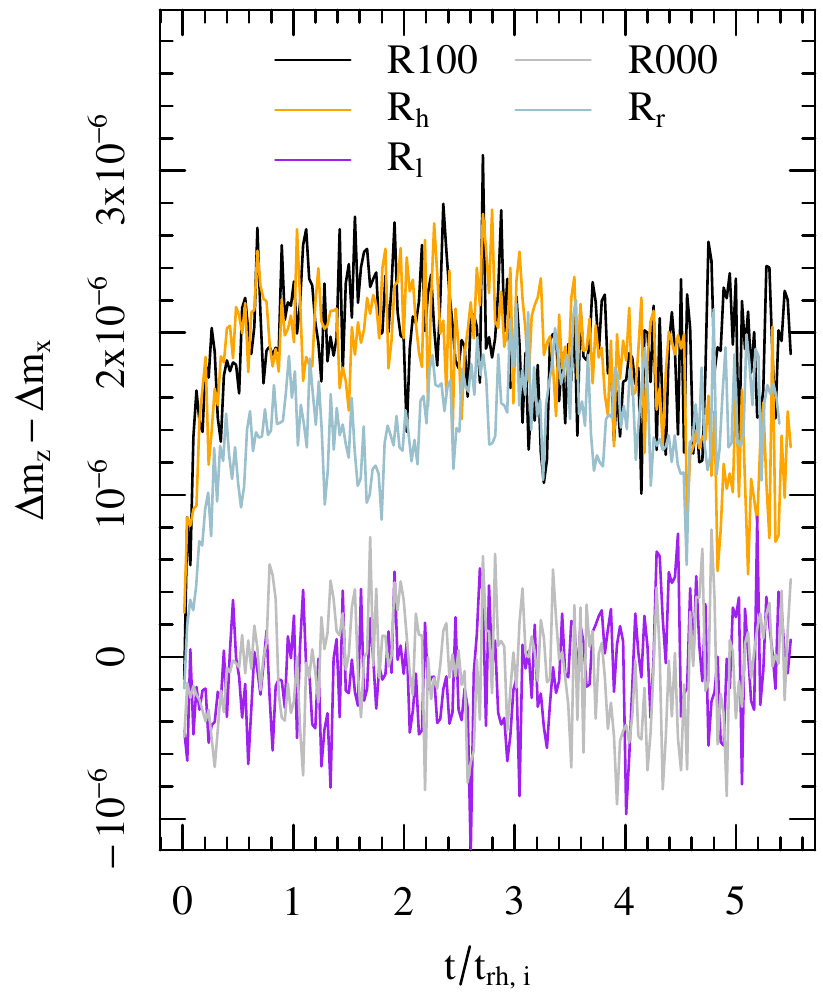}}
\caption{Evolution of $\deltam$ (see Eq. \ref{deltam} for definition) $z$-axis minus $\deltam$ along the $x$-axis, comparing the model where only the heavy population is rotated with the model where the light population is rotated.  Also shown is the model $R_{\rm r}$ that begins with a more realistic rotation curve.  Models $R100$ and $R000$ are shown for comparison.}
\label{fig:rotcompare2}
\end{figure}

However, the results plotted in Fig. \ref{fig:rotcompare2} show that the effect found in $R_h$ is identical to the effect measured in $R100$ for the first few relaxation times. This shows the importance of the density distribution of the massive rotating core. From an additional simple test performed with an $N$-body model of a rotating cluster without a mass spectrum, we noted that velocity anisotropy alone is not enough to induce this effect. From the $R_h$ model we learn that a massive rotating core produces a large pressure gradient, allowing for a much more prominent flattening than what it is observed in $R000$ and $R050$ models, which have similar velocity anisotropy but less rotation attributed to massive particles. At late times, the $R_h$ model shows a brief departure from the stable value seen in $R100$. This is due to the different characteristics of the models. At late times, $R100$ has a flattened, rotating, massive core but the light stars in the outer shells are also rotating. $R_h$, on the other hand, has a flattened, rotating, massive core, but its outer shells are isotropic and non-rotating. The massive core in the $R_h$ model is exchanging angular momentum with the non-rotating lower mass particles, which, in turn, reduce the flattening effect slightly; this does not happen in the $R100$ model. 

For comparison, we present the complementary model with the lower $2/3$rds of massive particles rotating in the same direction and the most massive particles remaining isotropic. Fig. \ref{fig:ics} shows the initial velocity anisotropy parameter of this model labeled as $R_l$.  The $\beta$ profile of this model is most similar to the $R075$ model. Again, if velocity anisotropy is the only predictor of the flattening effect, we would expect the $\Delta m$ measurement of $R_l$ to be similar to this model. We see in Fig. \ref{fig:rotcompare2} that this is not the case, and there is no measurable anisotropic mass segregation. This experiment can be simply explained: mass segregation taking place in the radial direction displaces the massive (and in this case, isotropic and non-rotating) particles to the core of the system. Without a rotating core, we do not see the steep pressure gradient around the axis of rotation. As the system expands, the low mass particles escape carrying away angular momentum which further reduces the spin of the system. By the end of the simulation, the core of the system remains isotropic and spherical.

\begin{figure}
\centerline{
 \includegraphics{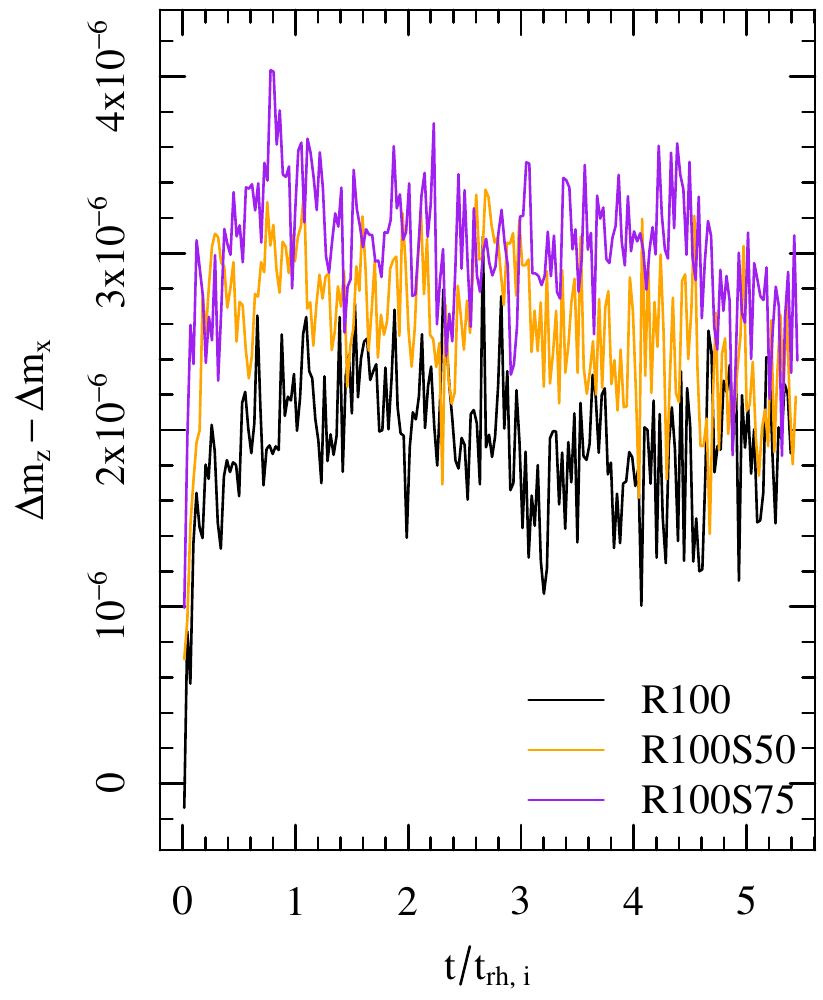}}
\caption{Evolution of $\deltam$ (see Eq. \ref{deltam} for definition) $z$-axis minus $\deltam$ along the $x$-axis, comparing the series of models where cluster begins nearly fully mass segregated ($R100S075$), slightly less mass segregated ($R100S50$), and no initial mass segregation (R100), and each cluster begins fully rotated.}
\label{fig:rotcompare3}
\end{figure}

\subsubsection{Experiment 2: Initially Radially Segregated Clusters}
The second set of numerical experiments involve initial conditions that are characterised by primordial isotropic mass segregation (see Section 2 for a description of the procedure adopted). With a fraction of massive particles initially located in the spherical core, we can determine how mass segregation in the radial direction affects the flattening of the massive core. We present the results from two $N$-body models. The initial conditions of model $R100S50$ are $50\%$ radially segregated, with all particles rotating in the same direction. Similarly, model $R100S75$ is $75\%$ radially segregated, with all particles rotating in the same direction.  The mass segregation parameter of interest, $\Delta m$, for both models is illustrated in Fig. \ref{fig:rotcompare3}, together with the result from $R100$ for comparison. 

We find that $N$-body models with some isotropic primordial mass segregation develop flatter cores (i.e., higher $\deltamz-\deltamx$), implying that systems that are initially more radially segregated can also reach higher levels of anisotropic mass segregation. Stated more directly, the simultaneous processes of radial mass segregation and flattening lead to less oblateness of the core when compared to a model that is already partially radially mass segregated.  This is expected according to the dynamical interpretation we have presented so far: in the initially non-segregated models, heavy particles in the outer regions must lose angular momentum to move toward the core via mass segregation in the radial direction. This loss of angular momentum affects particularly the central massive rotating core. On the other hand, in the initially segregated models the heavy particles in the centre maintain their angular momentum, therefore increasing the pressure gradient and allowing for further flattening in comparison. 

\subsubsection{Experiment 3: Realistic Rotation Curve}
\begin{figure}
\centerline{
 \includegraphics[width=0.5\textwidth] {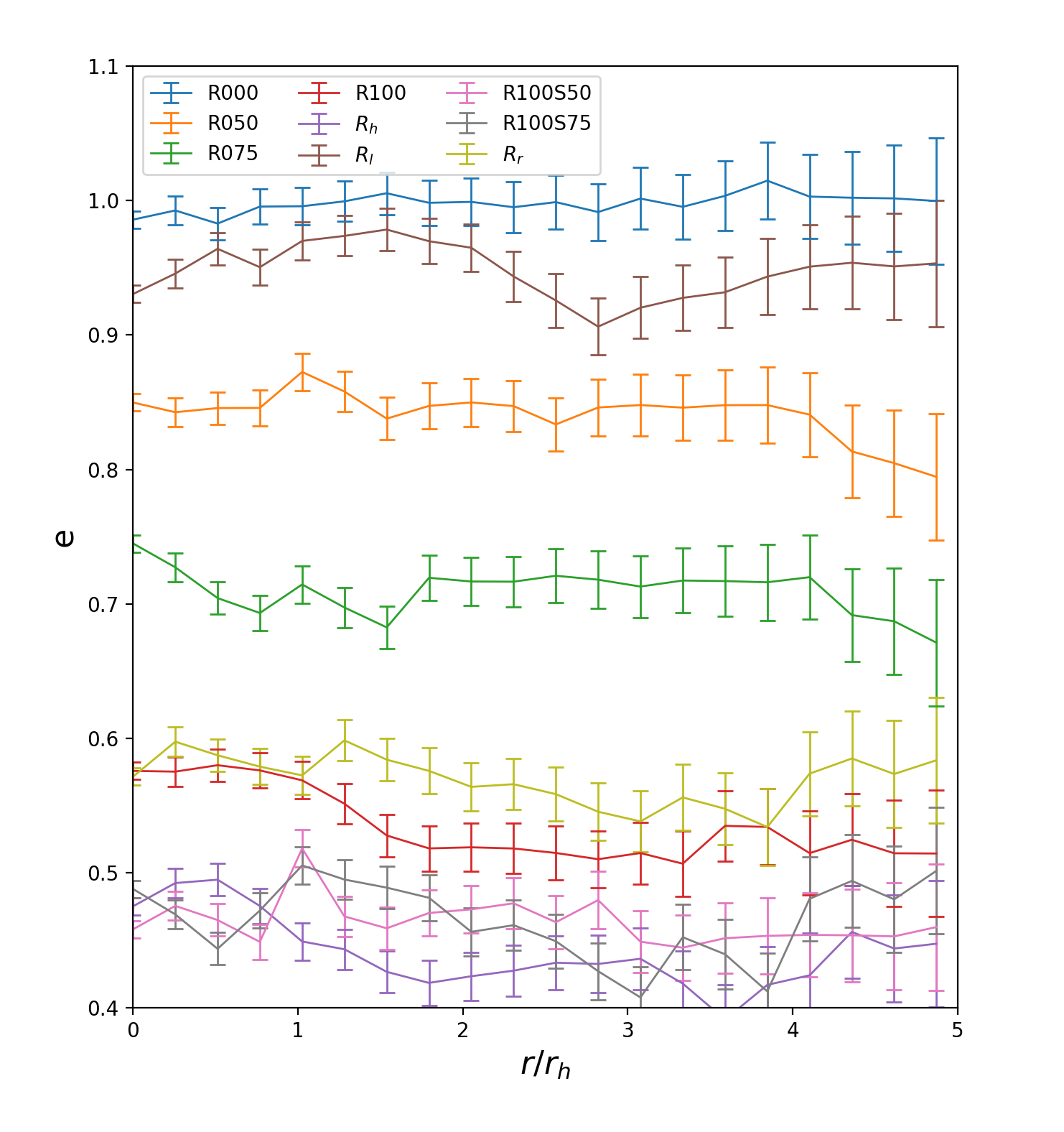}}
\caption{Ellipticity (defined as the ratio between the minor and the major axis) of the projected isodensity contours of all models and experiments in the $(x,z)$ plane at $t=1.38 \trhi$. The error bars represent the standard error of the mean for each bin.}
\label{fig:ellipticity2}
\end{figure}

The final numerical experiment shows the evolution of a system with a more realistic initial rotation curve (see Section 2 for a description of how this model was initialized). By differing the fraction of rotated particles at each radius, we can ensure that the flattening effect will persist in systems where the rotation curve varies with radius. We present the evolution of this $N$-body model, labeled $R_r$, in Figs. \ref{fig:rotcompare2} and \ref{fig:ellipticity2}, the latter of which we compare the ellipticity  profiles of all the models presented in this paper. We find that the system flattens significantly, most closely resembling the $R100$ model, although not as flat due to having less rotation in model $R100$ at the centre. This result matched our expectations from Experiments 1 and 2; the introduction of rotating massive particles and velocity anisotropy will produce flattening of the cluster during its evolution.

Going back to Fig. \ref{fig:ellipticity2}, we find that the order of all of our models according to ellipticity is mostly consistent with the order of the models measured with the anisotropic mass segregation parameter: with the fiducial models becoming more flat as more rotation is added, and the initially partially mass segregated models achieving higher levels of anisotropic mass segregation.  The exception appears to be model $R_h$ which has similar levels of aniostropic mass segregation as $R100$ but a flatter ellipticity profile.

\section{Conclusions}
\label{sec:conclusion}
The goal of this study was to investigate the connection between the morphological, structural and kinematical evolution of the central regions of collisional, multi-mass stellar systems, with special focus on the spatial characteristics of the process of mass segregation. 
Below, we briefly summarise the main findings of this exploratory work. 

First, when compared to non-rotating configurations with otherwise identical initial properties, multi-mass rotating systems quickly form a long lasting oblate, spheroidal massive core. This feature appears to preserve itself in the system for several relaxation times. Although our $N$-body simulations are highly idealised, we suggest that such an oblate core could exist in present-day Galactic globular clusters. We recall that our initial conditions are, by construction, spherical, but we propose that this morphological development will take place also in initially prolate or oblate systems, with an expectation for a stronger effect in the latter case, as an initially oblate system will have more massive particles near the core due to its geometry. Following the results from our additional numerical experiments on $N$-body models with primordial mass segregation, we would expect any oblate geometry to enhance the flattening effect.

Second, the degree of flattening experienced by the system is directly proportional to the initial degree of internal rotation. Observational measurements of an oblate core in nearby globular clusters could therefore hint at angular momentum being present in the system, if not now, then at some stage in the past. Given the persistence of such a feature on relatively long time scales, an observed oblate spheroidal core can therefore also provide a useful probe of the previous kinematic history of a stellar system, with possible implications for a number of on-going observational (e.g, see \citealt{Kamann2019, Trevino2019}) and theoretical (e.g, see \citealt{Lahen2020, Ballone2020}) efforts devoted to the study of the dynamical properties of young star clusters and star forming regions, for which a growing level of kinematic complexity is emerging.
 
Third, the flattening effect has a clear characterisation in terms of orbital architecture: it lowers the inclination of the orbits of massive stars. Such a behaviour could therefore have important ramifications on the spatial distribution of dark remnants in globular clusters and other dense stellar systems in the presence of internal rotation  (e.g., see \citealt{Webb2019,sz2019,Gruzinov2020}), with special reference to stellar-mass black holes and neutron stars that are currently at the centre of much attention given their role as possible sources of gravitational waves.    

In summary, we hope that this contribution can stimulate some renewed attention to be devoted to the morphology of star clusters, as their structural properties offer an essential counterpart to any kinematic characterisation and, therefore, represent a critical tool to fully reconstruct the phase space evolution of this class of stellar systems.

\section*{Acknowledgements}
Many thanks to Ann-Marie Madigan and the Eccentric Dynamics research group at JILA for helpful feedback and  discussion. 
This work utilized the Big Red II Supercomputer at Indiana University, which is supported in part by Lilly Endowment, Inc., through its support for the Indiana University Pervasive Technology Institute, and in part by the Indiana METACyt Initiative. The Indiana METACyt Initiative at IU was also supported in part by Lilly Endowment, Inc.  This work also utilized resources from the University of Colorado Boulder Research Computing Group, which is supported by the National Science Foundation (awards ACI-1532235 and ACI-1532236), the University of Colorado Boulder, and Colorado State University. ALV acknowledges support from a  UKRI Future Leaders Fellowship (MR/S018859/1).

\section*{Data Availability}

The data underlying this article will be shared on reasonable request to the corresponding author.

\bibliographystyle{mn}
\bibliography{main}

\appendix
\section{Evolution shortly after $t=0$}
We include here the same kinematical properties as a function of radius as in Fig. \ref{fig:ics}, but at $t=30$ H\'enon units, where some evolution can take place, but not long enough that two-body relaxation effects start to occur.  The rotational velocity profiles of all the rotating models evolve to resemble the characteristic rising, peaking, and falling rotation curves seen in observations of globular clusters.

\begin{figure*}
\centerline{
\includegraphics{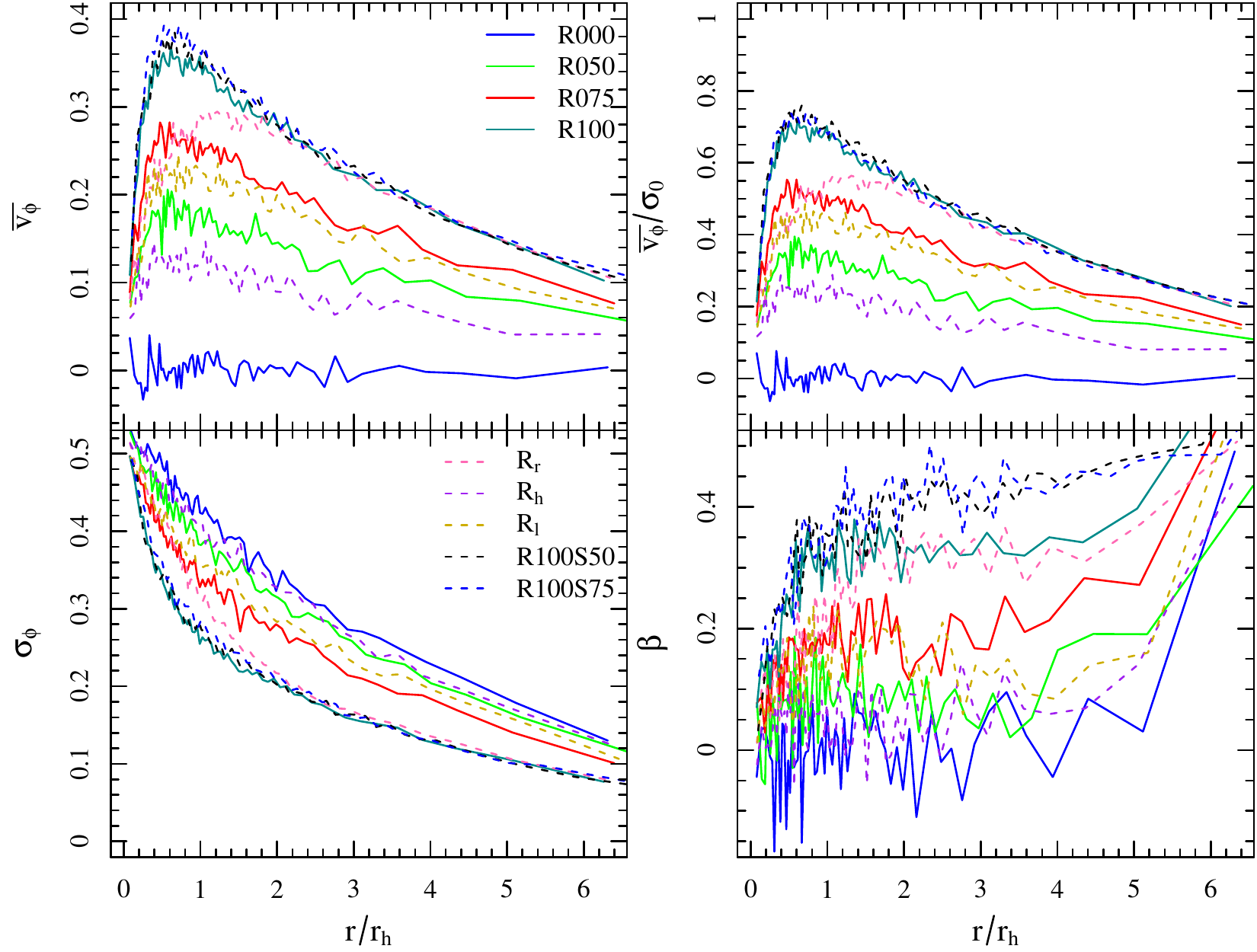}}
\caption{The same kinematical properties as a function of radius as Fig. \ref{fig:ics}, but at $t=30$ H\'enon units.}
\label{fig:t30}
\end{figure*}


\end{document}